\documentclass[fleqn,usenatbib]{mnras}

\usepackage{newtxtext,newtxmath}

\usepackage[T1]{fontenc}

\DeclareRobustCommand{\VAN}[3]{#2}
\let\VANthebibliography\thebibliography
\def\thebibliography{\DeclareRobustCommand{\VAN}[3]{##3}\VANthebibliography}

\usepackage{graphicx}	
\usepackage{amsmath}	
\usepackage{threeparttable}

\def \NYVir{NY\ Vir\ }

\def \Mjup{ M\textsubscript{Jup}\ }

\title[Orbital Stability of NY Vir Exoplanets]{Orbital Stability of Proposed NY Virginis Exoplanets}

\author[Mai \& Mutel]{
Xinyu Mai,$^{1}$\thanks{E-mail: xinyu-mai@ku.edu}
Robert L. Mutel$^{1}$\thanks{E-mail: robert-mutel@uiowa.edu}\\
$^{1}$Department of Physics and Astronomy, University of Iowa, Iowa City, IA 52242, USA\\
}

\date{Accepted XXX. Received YYY; in original form ZZZ}

\pubyear{2022}

\begin{document}
\label{firstpage}
\pagerange{\pageref{firstpage}--\pageref{lastpage}}
\maketitle

\begin{abstract}

Er et al. (2021) recently proposed a two-planet solution to account for eclipse timing variations (ETVs) observed from the sdB binary NY Virginis. We tested the proposed planetary system for orbit stability using both numerical simulations and chaotic behavior analysis. The best-fit orbits, as well as those with parameters varying by the published uncertainty range in each parameter, were unstable on a timescale much less than the presumed lifetime of the PCEB phase ($\sim$100 Myr). suggesting that the proposed circumbinary companions fail to provide a complete explanation for the observed ETVs. 

\end{abstract}




\section{Introduction}
Eclipse timing variations (ETVs) have been used to infer the existence of orbiting sub-stellar companions in more than a dozen short-period binaries, the majority of which are post-common binaries with hot sub-luminous primaries with orbital periods 2 -- 3 hours \citep{Heber:2016}. Since the substellar components perturb the center of the mass of the binary, their masses and orbital elements can be inferred by fitting the O-C plot, i.e., the time difference history between the observed times of eclipse minima and a linear ephemeris, ideally over a timescale longer than the orbital period of the most distant companion.   

This scheme has proven to be problematic for several reasons. First, typical timing offsets caused by a Jovian size planet are relatively small, of order one percent of the orbital period i.e., tens of seconds. This is not much larger than many published timing uncertainties, which are obtained using small telescopes that typically perform  synoptic observations of these systems. The corresponding fitted model parameters have large uncertainties and/or degeneracies.  Second, several other physical mechanisms can contribute to timing variations in these binaries (e.g., angular momentum transfer, apsidal motion, relativistic gravitational radiation), making even the existence of substellar components somewhat conjectural \citep{Applegate:1992,Lanza:1998,Zorotovic:2013,Lanza:2020,Mai:2022, Pulley:2022}. Third, several published multi-component models have  been subsequently shown to be highly unstable on timescales that are short compared with the expected lifetimes of the post-common envelope phase, vitiating the viability of the respective sub-component models.  \citep[e.g.,][]{Horner:2012,Wittenmyer:2012,Wittenmyer:2013,Brown:2021,Esmer:2021}.

\begin{figure}
\centering
\includegraphics[width=86mm, scale=1]{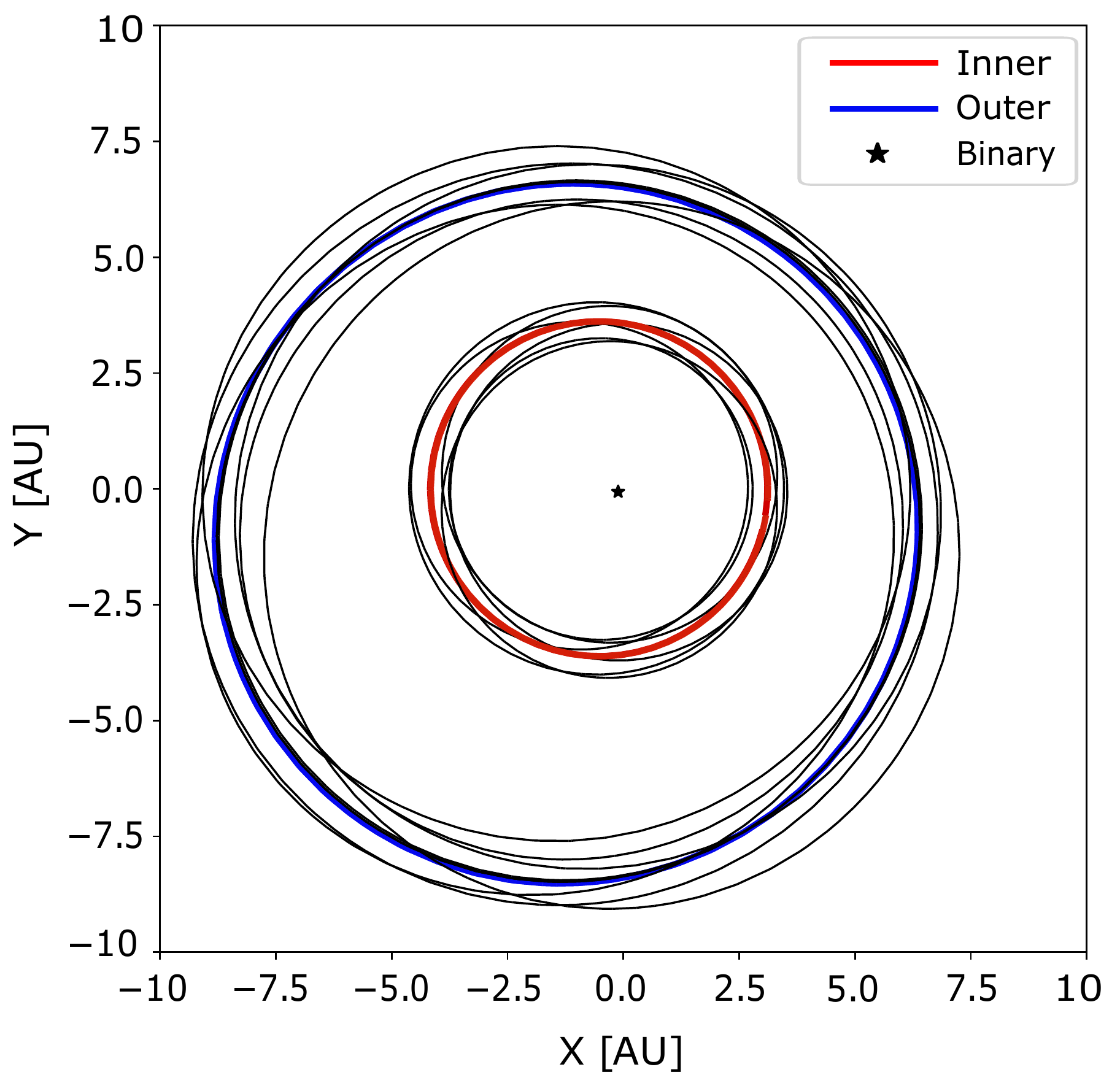}
\caption{Best-fit two-planet orbits proposed by \citet{Er:2021} around the PCEB \NYVir, and neighboring orbits using best-fit parameters, but varying by the uncertainty range given by Er et al. in each parameter. The trajectory in red indicates for proposed inner body, trajectory in blue for the outer body. Orbits associated with the change of one uncertainty are shown as black lines. The binary star position is shown as a star at the center. This figure is made by orbit simulation software $\textsc{rebound}$ \citep{Rein:2012} using best-fit parameters for each component in Table~ \ref{tab:params}. }
\label{fig:orbits}
\end{figure}

In this letter, we examine orbit stability of the two-component model recently reported by \citet{Er:2021} for the sdB binary NY Virginis. Their model is based on previously published O-C timing plus 51 new mid-eclipse times, extending the timeline to 25 years. Their model comprises two planets with masses of 2.74 \Mjup and 5.59 \Mjup with semi-major axes 3.64 and 7.64 AU and eccentricities of 0.12 and 0.19 respectively (see Table\ref{tab:params} for full orbital elements). Their derived planetary masses are similar to those of \citet{Song:2019}, but with a higher eccentricity for the outer planet. The best-fit orbits, along with neighboring orbits encompassing Er et al.'s uncertainty range in each parameter, are plotted in Fig.\ref{fig:orbits}.

\begin{table}
\centering
    \caption{\NYVir Two Component Model (\citealt{Er:2021})}
    \begin{threeparttable}[b]
    \label{tab:params}
    \centering
    \tabcolsep=0.2cm
    \renewcommand{\arraystretch}{1.3} 
    \begin{tabular}{r r r l}
        \hline
        \hline
        \multicolumn{1}{c}{Parameter} & \multicolumn{2}{c}{Fitted Values} & \multicolumn{1}{c}{Unit}\\
        \hline
        \multicolumn{4}{c}{Inner Binary} \\
        \hline 
        T$_0$     & \multicolumn{2}{l}{2453174.442636}          & BJD       	 \\
        P$_0$     & \multicolumn{2}{l}{0.1010159686}             & day	         \\
        $\dot{P}$  & \multicolumn{2}{l}{${2.64}$}	& $10^{-12}$\ s/s  \\
        \hline
        \multicolumn{4}{c}{ Substellar Components} 	\\
        \hline
        	         & LTT 1           	        & LTT 2	 	         &       \\
        $P$ 	     & $8.97_{-0.24}^{+0.36}$      &   $27.2_{-1.2}^{+1.3}$      & yr    \\
        $e$	         & $0.12_{-0.12}^{+0.12}$       & $0.19_{-0.07}^{+0.09}$     &       \\
        $K$          & $7.9_{-0.8}^{+1.0}$       & $33.8_{-2.9}^{+2.4}$     & sec	 \\
        $a \cdot$sin($i$)\tnote{1} 	 & $3.64_{-0.37}^{+0.41}$   & $7.64_{-0.43}^{+0.45}$         & AU	 \\
        $\omega$	 & $351_{-52}^{+69}$   & $38_{-34}^{+42}$  & degree \\
        Min. Mass	 & $2.74_{-0.34}^{+0.37}$  & $5.59_{-0.49}^{+0.51}$ & \Mjup  \\
        \hline
    \end{tabular} 
    
    \begin{tablenotes}[flushleft]
      \item[1] Planetary semi-major axes calculated from mass and period for each planet from the \citet{Er:2021} Table~3 using Kepler's third law.
    \end{tablenotes}
    \end{threeparttable}
\end{table}

\section{Orbit Stability Analysis}

To examine the orbital stability of the proposed planetary system, we performed dynamical simulation using the best-fit planetary masses and orbital elements listed in \citet{Er:2021} Table~3. In particular, we converted the projected semi-major axis of the binary star due to two additional companions $a_{1,2} \sin_{i_{3,4}}$ into planetary semi-major axes using the given mass and period for each planet. In addition, we tested  models for which each mass and orbital parameter were varied by the uncertainty ranges given in Table~3 of Er et al. from their best-fit value, for a total of 24 parameter sets. 

We used both the N-body orbital integration package REBOUND \citep{Rein:2012} and the chaotic indicator program, the Mean Exponential Growth factor of Nearby Orbits (MEGNO) \citep{Cincotta:2000, Hinse:2010}. We first tested dynamical stability by numerically integrating the component orbits using \uppercase{mercurius} \citep{Rein:2019}, a hybrid symplectic integrator  similar to MERCURY \citep{Chambers:1999}. It uses WHFast, a Wisdom-Holman symplectic integrator \citep{Rein:2015a} for long term integrations, but switches to IAS15, an adaptive non-symplectic integrator \citep{Rein:2015} for close encounters.

We used an initial time step of 0.1~yr and integrated for 100 Myr. We treated the binary star as a single mass $M_{tot} = 0.59 M_{\odot}$ \citet{Vulkovich:2009} and restricted the model to co-planar orbits. 
In addition, we tested chaotic behavior of each parameter set  using the MEGNO indicator with a time interval 10~Myr to search for chaotic phase-space regions. For a given integration time, MEGNO numerically evaluates a dimensionless quantitative measure of chaos in the system, and generates a mapping of phase-space regions with MEGNO values \citep{Gozdziewski:2001, Hinse:2010}. 
For the MEGNO calculations, we varied two orbital parameters, the semi-major axis and the eccentricity. We ran simulations for a range of 128 semi-major axes and 128 eccentricity values for the inner and outer planets. 

\section{Results}

\begin{figure}
\centering
\includegraphics[width=88mm,scale=1]{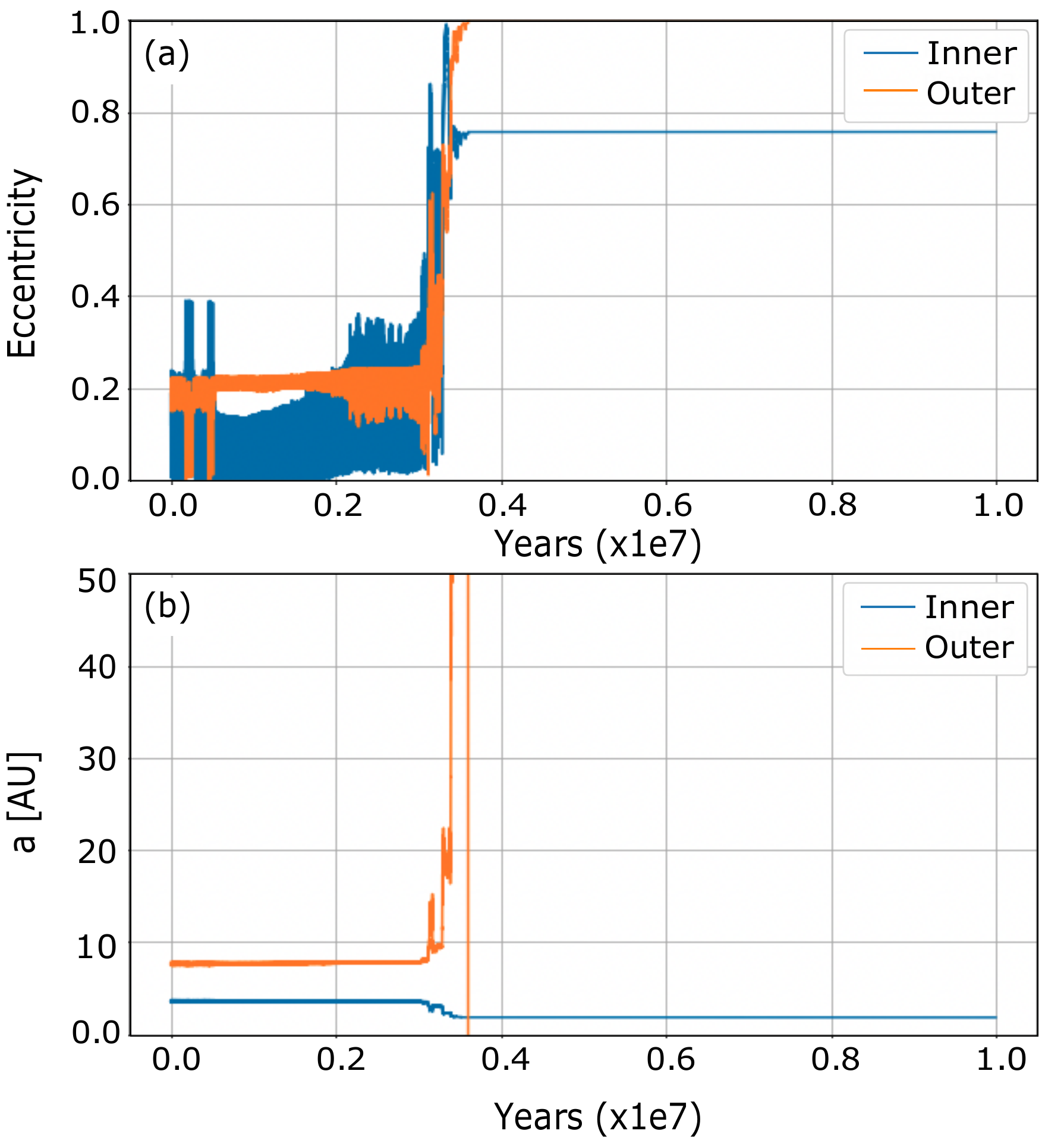}
\caption{NY Vir stability plot using the best-fit two-component planetary model of \citet{Er:2021}. (a) Time history of the eccentricities of both circumbinary companions over 10~Myr. (b) time history of semi-major axes of both circumbinary companions over 10~Myr. Note that the outer planet (orange line) escapes the planetary system after 3.5 Myr.}
\label{fig:rebound}
\end{figure}

\begin{figure*}
\centering
\includegraphics[angle=0,width=1.0\textwidth]{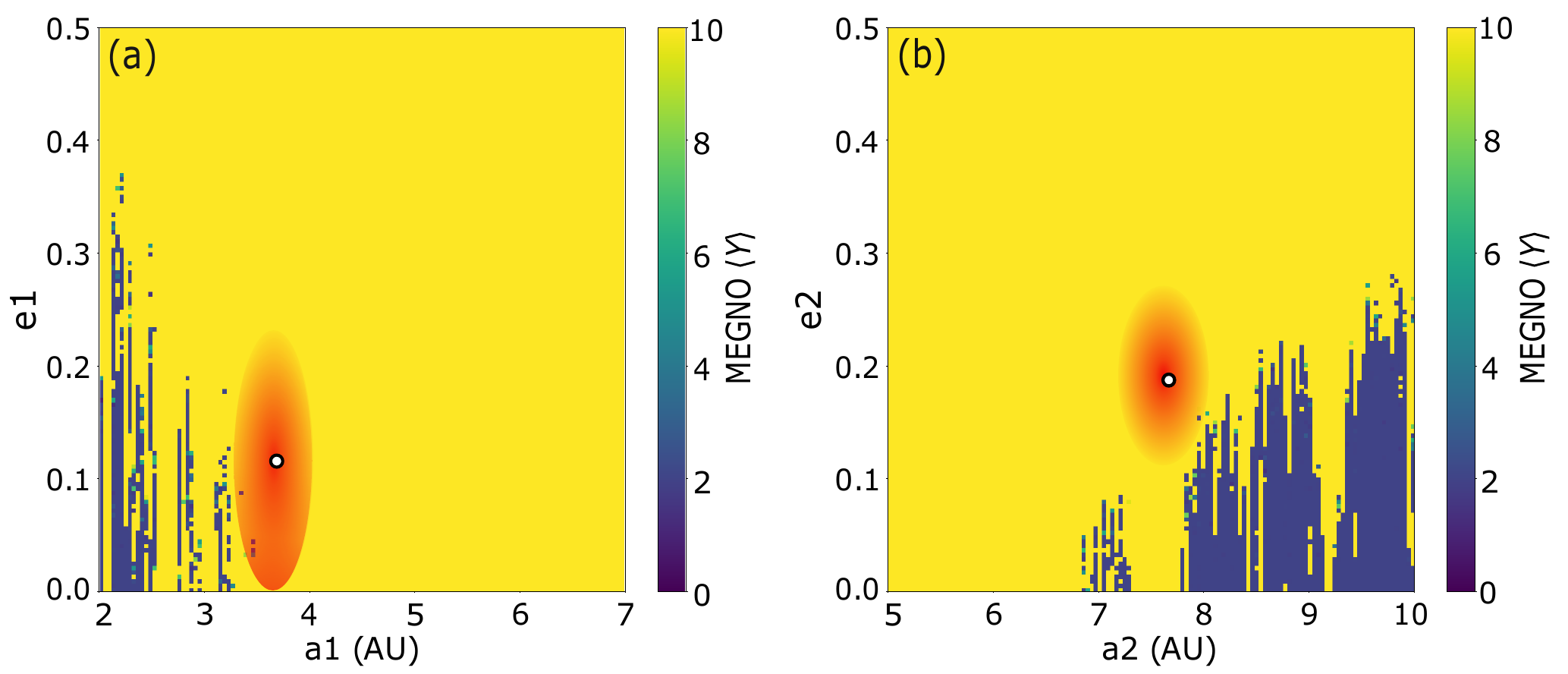}
\caption{MEGNO chaos parameter surface map for \protect\cite{Er:2021} two-planet solution for a range of eccentricities and semi-major axes for the (a) inner planet and (b) outer planet. The best-fit model parameter values (Table ~\ref{tab:params}) are indicated by white dots with black borders. Both points lie in highly chaotic regions. The shaded red regions are the approximate posterior probability distribution functions for eccentricity and semi-major axis of each planet based on \protect\cite{Er:2021} Figure 4.  ${\it Note}$: MEGNO values less than 2 are consistent with non-chaotic behavior \protect\citep[e.g.,][]{Hinse:2010}. }
\label{fig:megno}
\end{figure*}

As shown in Figure~\ref{fig:rebound}, our simulated dynamical result shows that the proposed best-fit planetary system of Er et al. becomes unstable for evolutionary times longer than $\sim$3~Myr, which is much less than the timescale of the PCEB phase for second-generation planets. In addition, the MEGNO chaos parameter, shown in Fig.~\ref{fig:megno}, is greater than 10 at the proposed solution coordinates (white dots) for both planets, implying highly chaotic behavior (N.B. values less than 2 imply non-chaotic behavior,  \citep[e.g.,][]{Hinse:2010}). 

\section{Discussion and Conclusion}
 We find that the proposed two-planet model for the binary NY Virginis proposed by \cite{Er:2021} becomes unstable after $\sim$3~Myr, a timescale much less than the estimated lifetime of the PCEB phase (100 Myr) of sdB binaries, and is therefore untenable. In addition, we find that all solutions that vary from the best-fit solution by the uncertainty range in each parameter are also unstable on similar or shorter timescales. This conclusion is supported by an independent stability indicator, the chaos metric MEGNO.  We evaluated the MEGNO  map chaotic behavior in parameter space over 10~Myr and found that both planets have orbital parameters than lie in regions of highly chaotic behavior.

The previous two-planet solution for NY Vir proposed by \cite{Song:2019}, although stable over more than 10~Myr, does not fit  more recent O-C data, as shown both by \cite[][Fig.3]{Er:2021} and \cite[][Fig.2]{Pulley:2022}. 
Hence, the situation for NY Vir is very similar to at least two other sdB binaries, HW Vir and HS0705+67: Sub-stellar component models that provide an adequate fit to the ETV historical data are dynamically unstable. Likewise, solutions that are constrained to be long-term stable do not result in adequate fits to the ETV data, as we have recently demonstrated \citep{Mai:2022}. 

As the number of these confounding systems increases, it is becoming clear that interpreting ETV variations in sdB binaries as resulting solely from gravitational perturbations from orbiting sub-stellar objects is untenable, and that the observed ETV signatures must be at least in part a result of other physical mechanisms.

\section*{Acknowledgements}

This research made use of $\textsc{astropy}$, a community-developed core $\textsc{python}$ package for Astronomy \citep{astropy:2013, astropy:2018}. Numerical orbit integrations in this paper made use of the $\textsc{rebound}$ N-body code \citep{Rein:2012}. The simulations were integrated using $\textsc{whfast}$, a Wisdom-Holman symplectic integrator \citep{Rein:2015a}. This work is supported by the National Science Foundation (NSF) under grant 1517412.

\section*{Data Availability}

There are no new data associated with this letter.



\bibliographystyle{mnras}
\bibliography{nyvir-library}








\bsp	
\label{lastpage}
\end{document}